\documentclass[journal=jctc,manuscript=article,layout=twocolumn]{achemso}
 
\usepackage{chemformula} % Formula subscripts using \ch{}
\usepackage[T1]{fontenc} % Use modern font encodings
\usepackage{amsfonts}

\DeclareMathOperator*{\argmax}{argmax}

\author{Pablo Herrera-Nieto}
\altaffiliation{These authors contributed equally to this work}
\author{Adri\`a P\'erez }
\affiliation{Computational Science Laboratory, Universitat Pompeu Fabra, Barcelona Biomedical Research Park (PRBB), C Dr. Aiguader 88, 08003, Barcelona, Spain}
\alsoaffiliation{Acellera Labs, C Dr Trueta 183, 08005, Barcelona, Spain}
\altaffiliation{These authors contributed equally to this work}
\author{Gianni De Fabritiis}
\email{gianni.defabritiis@upf.edu}
\affiliation{Computational Science Laboratory, Universitat Pompeu Fabra, Barcelona Biomedical Research Park (PRBB), C Dr. Aiguader 88, 08003, Barcelona, Spain}
\alsoaffiliation{Acellera Ltd, Devonshire House 582,
HA7 1JS, United Kingdom}
\alsoaffiliation{Instituci\'o Catalana de Recerca i Estudis Avan\c{c}ats (ICREA), Passeig Lluis Companys 23, 08010 Barcelona, Spain}

\title[Binding-and-folding KIX-cMyb]
  {Binding-and-folding recognition of an intrinsically disordered protein using online learning molecular dynamics}

\abbreviations{MD,MSM,IDP,UCB,NMR}

\begin{document}

\begin{abstract}
Intrinsically disordered proteins participate in many biological processes by folding upon binding with other proteins. However,
coupled folding and binding processes are not well understood from an atomistic point of view.
One of the main questions is whether folding occurs prior to or after binding. 
Here we use a novel unbiased high-throughput adaptive sampling approach to reconstruct the binding and folding between the disordered transactivation domain of \mbox{c-Myb} and the KIX domain of the CREB-binding protein.
The reconstructed long-term dynamical process highlights the binding of a short stretch of amino acids on \mbox{c-Myb} as a folded $\alpha$-helix.
Leucine residues, specially Leu298 to Leu302, establish initial native contacts that prime the binding and folding of the rest of the peptide, with a  mixture of conformational selection on the N-terminal region with an induced fit of the C-terminal.
\end{abstract}

\section{Introduction}
Intrinsically disordered proteins (IDPs) participate in many biological functions despite lacking a stable tertiary structure \cite{dyson2005intrinsically}. 
Initial clues for the function of IDPs were revealed by structural studies \cite{kussie1996structure,zor2004solution}, showing that proteins that were disordered in isolation became folded upon interacting with their partners, opening to question how folding couples with binding. 

Recently, molecular dynamics (MD) simulations have been successfully applied to reconstruct biological dynamic events in problems such as protein-ligand \cite{buch2011complete} and protein-protein \cite{plattner2017complete,borgia2018extreme} binding, as well as protein folding \cite{lindorff2011fast,piana2013atomistic}.
MD has also been applied in the field of IDPs \cite{zwier2016efficient,morrone2017computed,zhou2017bridging,paul2017protein}.
In particular, the Mdm2 protein and the disordered 12-residue N-terminal region of p53 were studied using implicit solvent simulations\cite{zwier2016efficient}, parallel full-atom simulations totaling \cite{zhou2017bridging}, biased free-energy-based sampling \cite{morrone2017computed}, and biased/unbiased simulations to estimate kinetics on the second timescale \cite{paul2017protein}.
For another system, KIX-pKID, a single event of binding \cite{chong2019explicit} has been sampled at all-atom resolution. 

\begin{figure*}[!ht]
\centering
\includegraphics[width=17.8cm, keepaspectratio]{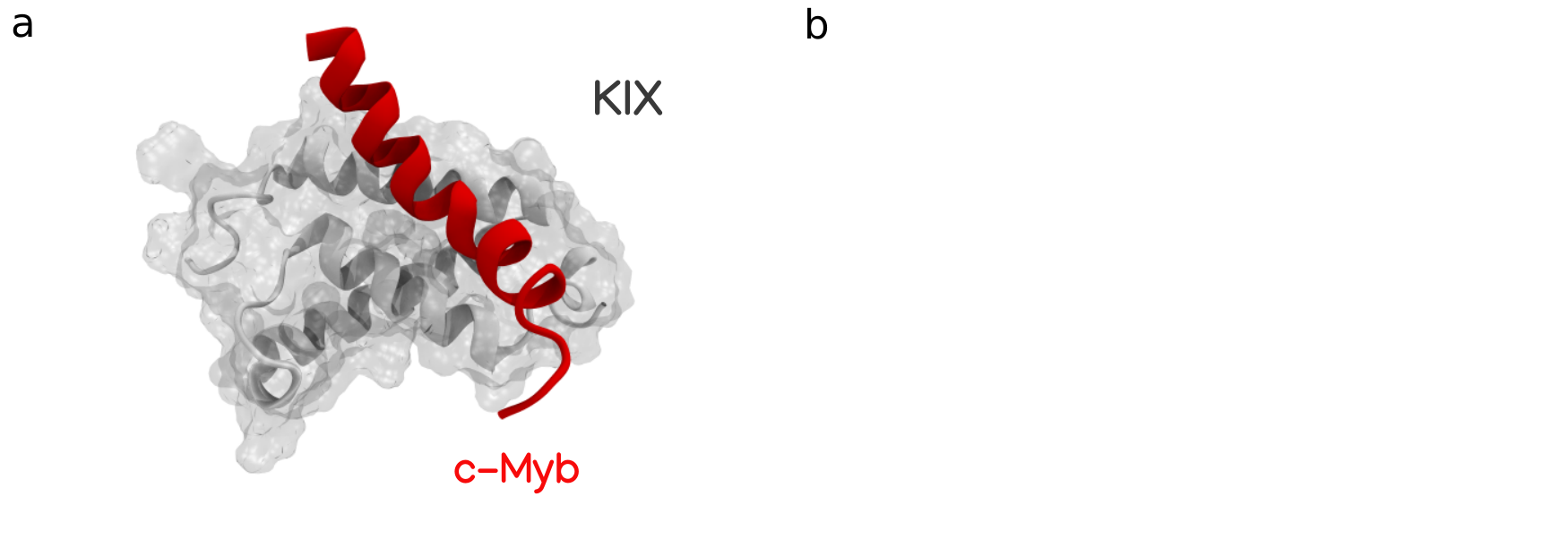}

\caption{\textbf{Exploration performance.} 
\textbf{a)} KIX---cMyb NMR structure. KIX domain is shown as a white surface and ribbon and c-Myb bound to KIX as a red helix (PDB code \textit{1SB0}).
\textbf{Exploration performance} by 
\textbf{b) Counts Adaptive} ($\sim480\ \mu s$)
, and
\textbf{AdaptiveBandit}  ($\sim450\ \mu s$)
is shown by plotting the mean RMSD (on the x-axis) and standard deviation (on the y-axis) for each of the MSM's microstates, color-mapped accordingly to their macrostate assignment.
The dashed square indicates the \textit{bound zone}, placed in the region corresponding to low mean RMSD and standard deviation.
}
\label{fig:sampling}
\end{figure*}

The \mbox{KIX---c-Myb} binding-and-folding mechanism has been extensively studied experimentally as an exemplar case of protein-IDP interaction \cite{arai2015conformational,giri2013structure,gianni2012folding,shammas2013remarkably,toto2016molecular,poosapati2018uncoupling,shammas2014allostery}. 
The KIX domain of the CREB-binding protein is a short 87-aa region composed of three $\alpha$-helices (designated as $\alpha$-1, $\alpha$-2 and $\alpha$-3, from N-terminal to C-terminal) forming a compact bundle \cite{zor2004solution}. 
KIX represents a paradigm of binding promiscuity: it binds to many IDPs, including the proto-oncogene \mbox{c-Myb} \cite{zor2004solution} (Figure \ref{fig:sampling}.a), with multiple binding conformations \cite{arai2015conformational}.
However, the system composed by \mbox{KIX---c-Myb} remained outside of the scope of all-atom molecular simulations due to the size of the IDP (it doubles the length of p53) and the existence of multiple binding modes between them \cite{arai2015conformational}. 
In particular, it is unclear whether the interaction takes place by conformational selection, i.e. \mbox{c-Myb} needs to be folded before binding to its partner, or by \mbox{induced-fit}, where  binding not only happens independently of \mbox{c-Myb}'s secondary structure but also triggers its folding, as shown for other IDPs (KIX-pKID) \cite{sugase2007mechanism,chong2019explicit}. Understanding these aspects has implications for the druggability of disordered proteins.
Another important factor is \mbox{c-Myb}'s high helicity in isolation and the consequences it might exert on the final complex structure, which features an extended $\alpha$-helical \mbox{c-Myb} bound to KIX.
Some reports support the \mbox{induced-fit} approach based on kinetics and mutagenesis studies \cite{gianni2012folding,giri2013structure,shammas2014allostery}, while others advocate for a mixed mechanism \citep{arai2015conformational}; yet not a detailed model for the binding process is available.

In this paper, we take advantage of a novel algorithm that frames the MD sampling problem from a reinforcement learning perspective (see \cite{perez2020adaptivebandit} and Methods) to reconstruct multiple binding modes between \mbox{c-Myb} and KIX.
This sampling algorithm was key for us to reconstruct the binding process, as previous attempts over the years using other state-of-the-art adaptive sampling methods \cite{doerr2014fly,doerr2016htmd} were not successful, always failing to recover the NMR bound structure.
Results provide insights into the binding mechanism between these two proteins, supporting a mixed model that combines both conformational selection and induced fit.

\section{Results \& Discussion}

\subsection*{Adaptive sampling the KIX---c-Myb binding-and-folding process}
Simulations to reconstruct the \mbox{KIX---c-Myb} binding mode were performed following an adaptive sampling strategy.
In adaptive sampling, successive rounds of simulations are performed in an iterative step-wise manner, where an acquisition function over the currently sampled conformations is defined. 
We compare two of the acquisition functions used for \mbox{KIX---c-Myb} simulations: a count-based one and another one inspired by reinforcement learning, part of the novel AdaptiveBandit method \cite{perez2020adaptivebandit}.

The new AdaptiveBandit method is framed into a simplified reinforcement learning problem, the multi-armed bandit problem (see \textit{Methods}). We use the upper confidence bound (UCB) algorithm \cite{auer2002using} to optimize an action-picking policy in order to maximize future rewards, optimally balancing the exploration of new higher rewarding actions with the exploitation of the most known rewarding ones. The reward function, which associates the action with the reward given by the system, defines what we want to optimize. In this work, we choose the reward to be minus the free energy of each configuration visited in the trajectory spawn from a given action (see Eq.\ref{reweq} in \textit{Methods}), where the free energy of a conformation is given by the corresponding Markov state model (MSM) microstate computed with the data available at the current sampling epoch. 

Standard low counts adaptive sampling \cite{doerr2014fly} (hereby named Counts Adaptive) can be shown to be optimal in pure exploration conditions \cite{doerr2016htmd}. 
Counts are computed over clusters of conformations; this method is, however, noisy as clusters can be poorly populated. Therefore, in the implementation available in \cite{doerr2016htmd}, counts are computed over a smaller subset by grouping clusters (microstates) into macrostates, constructing a Markov State Model (MSM) \cite{prinz2011markov} with the available data at each round. 
The acquisition function is given by proportionally choosing macrostates as $1/c$, where $c$ represents macrostate counts, and by randomly selecting conformations within them. 

A comparison between Counts Adaptive and AdaptiveBandit is provided in Figure \ref{fig:sampling}.b. 
The batch based on Counts Adaptive (48 epochs) failed to connect microstates similar to the NMR structure in over $\sim480\ \mu s$, reaching at best an RMSD around 7 \AA.
For us, it was impossible to build an MSM with the bound state with previous methods, and novel approaches  were needed to reconstruct the binding-and-folding process between KIX and \mbox{c-Myb} successfully. 
AdaptiveBandit provides converged estimates of kinetics and thermodynamics after just $150\ \mu s$ of sampling (Supplementary Figure 5).

\begin{figure*}[!ht]
\centering
\includegraphics[width=\textwidth, keepaspectratio]{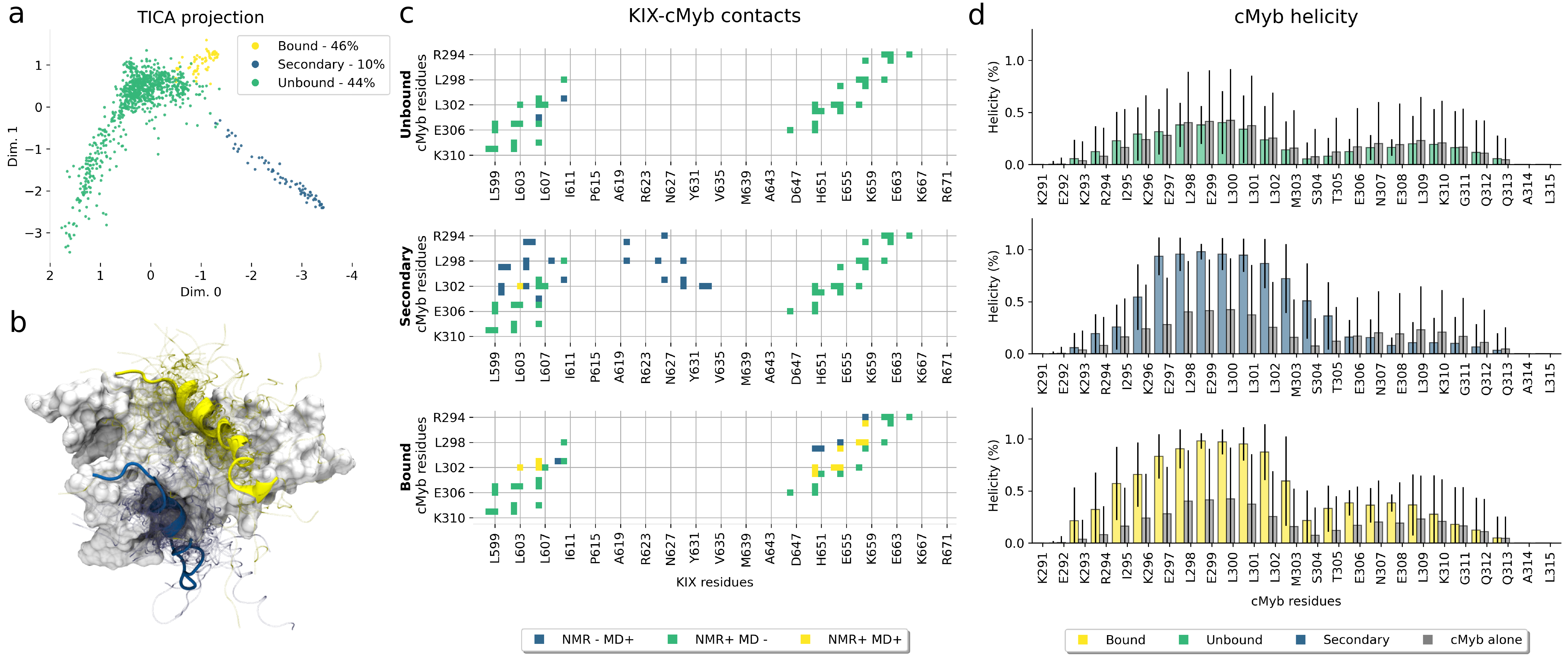}
\caption{\textbf{KIX \& c-Myb binding model.} 
\textbf{a) States distribution across the TICA space:} microstates are represented as dots and are colored following their macrostate assignment.  
\textbf{b) Representative structures:} PDB structure \textit{1SB0} is depicted with KIX as a gray surface, \mbox{c-Myb} bound to the primary interface as a yellow ribbon, and \mbox{c-Myb} bound to the secondary interface as a blue ribbon. c-Myb backbones for 30 representatives MD structures of \textit{bound} and \textit{secondary} states are displayed with blurry yellow and blue clouds, respectively.
\textbf{c) Macrostate contact fingerprint:} profile of contacts established between c-Myb and KIX in each macrostate in at least 50\% of the structures. 
Blue color represents contacts present in the state but not in the original NMR structure; green indicates original NMR contacts not found in the MSM state; and yellow squares represent contact matches, found in both NMR and MD structures.
\textbf{d) Macrostate cMyb helicity:} helicity fraction per residue of c-Myb in each macrostate. Helicity for the cMyb peptide alone is depicted in grey in each plot for comparison.
}
\label{fig:f1}
\end{figure*}

\subsection*{Identification of the bound state}
The full data set of the AdaptiveBandit run accounted for a total simulation time of $\sim$450 $\mu$s, split across 40 epochs, and was the one used to study the molecular features of KIX---c-Myb binding-and-folding.
For the analysis, a slightly different MSM was built based on all-pair $C_{\alpha}+C_{\beta}$ distances between KIX and \mbox{c-Myb}, self distances between $C_{\alpha}$ of \mbox{c-Myb}, secondary structure of \mbox{c-Myb} and RMSD to the NMR bound conformation (PDB ID: \textit{1SB0}). 
The MSM defines three kinetically similar sets of conformations, referred as macrostates (Figure \ref{fig:f1}.a and Supplementary Figure 1.b): a highly populated state with an heterogeneous mixture of conformations (\textit{unbound}), a well defined \mbox{c-Myb} bound state (\textit{bound}) and, finally, a secondary bound state (\mbox{\textit{secondary}}). 
Representative structures of all states can be found in Figure \ref{fig:f1}.b. The \textit{bound} macrostate contains structures with a minimum RMSD of approximately 3 {\AA} with respect to the NMR structure. Complete binding and folding trajectory videos can be found in Supplementary Table 1, with reconstructed trajectories from different epochs containing unique paths to the bound state (with RMSD $<4$ {\AA} with respect to the bound NMR structure)

The \mbox{\textit{bound}} state identifies the primary cMyb bound pose in the hydrophobic groove between $\alpha$-1 and $\alpha$-3) of KIX. On average, it shares 36\% of the fraction of native inter-molecular contacts ($Q_{int}$) with the original NMR structure, as shown in Figure \ref{fig:f1}.c.  
These contacts mainly involve the interaction of \mbox{c-Myb} residues Leu298 and Leu302 with residues across the primary binding interface: Leu302 contacts Leu603, Leu653, and specially Leu607 of KIX, which is buried down in the pocket, whereas Leu298 establishes additional native contacts with Ala610, Ile657, and Tyr658. 
$Q_{int}$ reaches up to 80\% in those microstates exhibiting the tightest bound conformations, and, in addition to the leucine binding, they feature most of the contacts between the C-terminal half of \mbox{c-Myb} and KIX, which are not that prevalent across the \textit{bound} macrostate (Supplementary Figure 2). 
The main contacts missing account for the electrostatic interactions established between Arg294 and the region on $\alpha$-3. There are some conformations where these interactions occur, but their prevalence in those microstates is less than 50\%.

The secondary structure profile for MD-derived states matches the experimental description of \mbox{c-Myb} \cite{arai2015conformational,poosapati2018uncoupling}, as shown in Supplementary Figure 3: the 25 residues are separated in two halves by residues Met303 and Ser304. 
The N-terminal half shows a high helical tendency, around 20-30\% for residues in positions 297 to 302 with \mbox{c-Myb} in isolation, 
being maximal in bound states. 
Experimentally, this N-terminal half in isolation reaches even higher helicity levels ($\sim$70\%) when using an extended construct of \mbox{c-Myb} \cite{arai2015conformational}. 
On the other hand, the C-terminal section exhibits low helical propensity when in isolation, and  increases when bound to KIX. The full helix conformation only appears in those microstates with the tightest bound conformations.

\begin{figure*}[!ht]
\centering
\includegraphics[width=\textwidth, keepaspectratio]{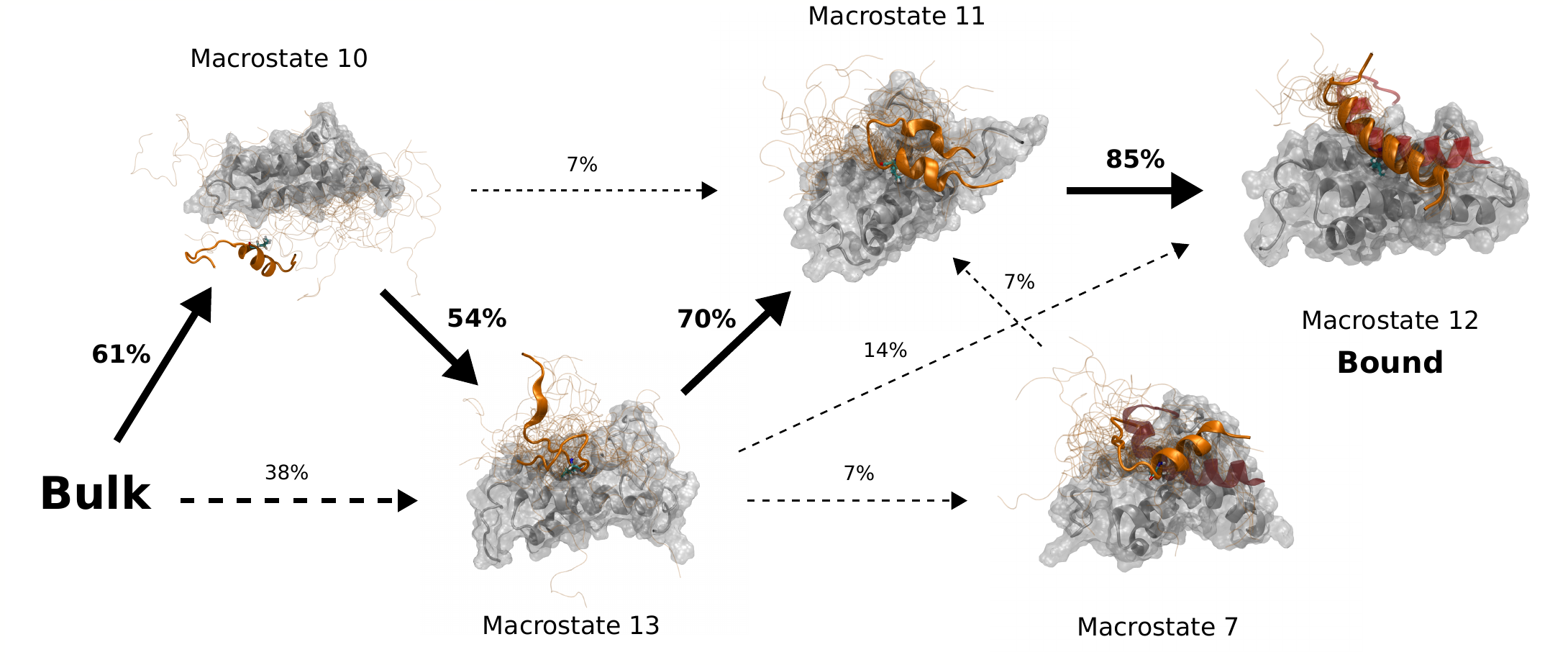}
\caption{\textbf{Complete c-Myb binding process to KIX domain.} 
Main pathways leading from \textit{Bulk} (macrostate 14) to the \textit{Bound} state (macrostate 12). Fluxes are shown as percentages near the arrows. Only those fluxes higher than 5\% are shown. The arrow thickness is proportional to the flux percentage. Straight arrows indicate the maximum flux path, while dashed arrows show other fluxes. Each macrostate structure shows KIX as the white surface and ribbons and c-Myb as the orange ribbon and tubes. For each macrostate, 25 conformations are shown as thin tubes with one structure highlighted as a ribbon structure that includes the side chain of Leu302, colored by atom element. Additionally, for macrostates 7 and 12, the reference NMR c-Myb structure is shown as a transparent red ribbon for comparison.}
\label{fig:binding}
\end{figure*}

\subsection*{Secondary binding mode}
The existence of alternative binding poses between \mbox{c-Myb} and KIX has also been reported \cite{arai2015conformational}.
The MSM shows the presence of a secondary binding mode (referred to as \textit{secondary}), occupying a novel interface, located between $\alpha$-1 and $\alpha$-2 (Figure 1.b and Supplementary Figure 9).
The interaction of the \textit{secondary} state resembles the \mbox{\textit{bound}} binding mode: the N-terminal half is folded in the typical $\alpha$-helix, while the C-terminal section remains mostly unstructured. The presence of a native contact in this secondary binding mode is due to the penetration of Leu302, located close to Leu603's backbone in KIX, rather than by side-chain proximity. Leu298 and Leu302 of \mbox{c-Myb} are deeply buried in a hydrophobic pocket composed of residues Val604, Val608, Leu620 (found in the G2 helix, which connects $\alpha$-1 and $\alpha$-2) and Val629.
Kinetically, there is a 10-fold difference in the mean first passage time for binding between both sites --- $(9.96\pm3.57)\cdot10^3$ ns for binding to \mbox{\textit{bound}} site and $(1.05 \pm 0.46)\cdot10^{5}$ ns for the \textit{secondary} site --- that may account for the preferential binding of \mbox{c-Myb} to the primary interface.

\subsection*{Model validation}
To validate the model, we compared the kinetic parameters derived from it with available information \cite{shammas2013remarkably}.
Experimental values from Shammas et al. were calculated at temperatures ranging from 278 to 298 K, while simulations were executed at physiological temperature (310 K).
$k_{on}$ values display a temperature independent tendency, whereas for temperature dependent variables $k_{off}$ and $k_{d}$ values had to be extrapolated to 310K (Supplementary Figure 4).
Hence, reference values for $k_{off}$ and free energy (obtained from $k_{d}$) resulted in 866 $s^{-1}$ and \mbox{-6.81 $kcal\ mol^{-1}$} respectively.

Due to the size of the peptide compared to the solvation box, it is hard for the MSM to automatically define the correct bulk state. We, therefore, manually defined a bulk state that contains conformations where the distance between KIX and cMyb is maximized. The bulk state was defined by taking those microstates where the minimum distance between KIX and cMyb is higher than a threshold.  Consequently, some kinetic and thermodynamic estimates have a dependency on such distance threshold (Supplementary Figure 6a) as this affects the definition of the bulk state. However, the computed $k_{off}$ and free energy estimates are practically stable after a minimum separation distance of just  4 {\AA}.

The obtained MSM estimations of $k_{on}$ go between \mbox{$(2.72)\cdot10^{7}M^{-1}s^{-1}$} and \mbox{$(3.65)\cdot10^{7}M^{-1}s^{-1}$}, in agreement with the experimental value $(2.2\pm0.2)\cdot10^{7}M^{-1}s^{-1}$ \cite{shammas2013remarkably}. $k_{off}$ estimates range from $3.50\cdot10^3\ s^{-1}$ to $21.70\cdot10^3\ s^{-1}$, overestimating the extrapolated experimental value by an order of magnitude. Free energy estimates range between $-7.35\ kcal\ mol^{-1}$ and $-6.27\ kcal\ mol^{-1}$ depending on the choices of the analysis parameters, with the extrapolated experimental value inside this interval. 

We further verified the reproducibility of kinetic and thermodynamic measurements  to ensure model convergence by building multiple MSMs using incrementally more trajectories. Convergence is reached at $150\ \mu s$ on all the previous estimates (Supplementary Figure 5). 
The discrepancy between the experimental reference and computed $k_{off}$ values translates into a faster dissociation in our model. We also verified if this was due to normal discretization errors in the MSM projection or to the fact that our simulations did not obtain a complete bound conformation between KIX and cMyb. In order to test this hypothesis, additional long trajectories (8 replicas of 2 $\mu$s each) were run starting from bound NMR and MD-derived conformations. We constructed an MSM using both simulation datasets. However, the free energy estimations are only marginally improved (Supplementary Figure 6b). Thus, we concluded that the additional bound simulations do not add additional information and we restrict  the analysis to just the AdaptiveBandit set of simulations as this is the most general case where no NMR information is available. 

\subsection*{Binding follows both \mbox{induced-fit} and conformational selection}
 In order to gain additional structural insight of the binding process, we constructed an MSM with a higher number of macrostates, using the same lag time.
We used  transition path theory \cite{weinan2006towards,noe2009constructing} to calculate fluxes leading from the purely bulk state to the bound conformations.
Out of the 15 macrostates of the new MSM, only a reduced set of 6 is sufficient to explain binding to the primary interface. The other macrostates describe either the secondary binding mode or other unstable interactions between KIX and cMyb.
The network generated by the flux interchanges between macrostates (Figure \ref{fig:binding}) separates the binding and folding process into three events: the establishment of the initial contacts, binding and folding of the N-term section of cMyb  and finally, binding and folding of the C-term section of cMyb. The first step of the binding process features the first native contacts found across the \mbox{KIX---c-Myb} binding pathway, which involves residues Leu302 of \mbox{c-Myb}. The role of Leu302 as the main driving force for the interaction has already been described \cite{zor2004solution} and is due in part to the kink in the helix created by neighbors residues Met303 and Ser304, which exposes Leu302 allowing for a deep penetration inside the binding pocket. Besides, of the KIX residues contacted at this stage is Leu603, which is one of the most exposed residues in the hydrophobic pocket later occupied by Leu302. 

To determine if cMyb folding precedes or follows binding at this step, we looked at the flux passing through different cMyb conformations in macrostate 13 (Supplementary Figure 8). We see that almost half of the flux goes through N-term helical conformations, suggesting that the presence of helix in residues 297 to 302 facilitates these first binding step, following a conformational selection mechanism. There is also a considerable flux going through unfolded conformations, meaning Leu302 also binds through induced fit. 

The second step of the binding process goes from macrostate 13 to macrostate 11, where several contacts are formed across the N-term of cMyb. The last step, which goes from macrostate 11 to 12 (the bound state), involves forming the last contacts on the C-term and completely folding cMyb. Here, we also looked at the flux passing through macrostate 11 (Supplementary Figure 8c) to discern between conformational selection and induced fit on the C-term folding and binding mechanism. Here, all the flux goes through conformations where the C-term is unfolded, meaning that only when the C-term native contacts start to happen we see a complete cMyb folding to an alpha helix, following a clear induced-fit mechanism.

The overall mechanism works as an induced-fit binding and folding, but we can see a mixed mechanism during the first binding steps, where both conformational selection and induced-fit seem to take a part in facilitating the first contacts through cMyb's Leu302.
In summary, initial steps can be greatly benefited from pre-folded helical structures of \mbox{c-Myb} (Figure \ref{fig:binding}), although \textit{binding before folding} is also observed.
The binding of helical conformations dominates the initial steps of the interaction, but for the interaction of the \mbox{C-terminal} tail, folding follows binding. No limiting steps in the binding process are observed; hence no possible transition states can be defined, as pointed out by experimental reports \cite{shammas2013remarkably}.

\section{Conclusions}

The analysis presented here provides a detailed molecular description of binding of \mbox{c-Myb} to the primary interface of KIX, summarized as a two-step process, where initially the N-terminal region of \mbox{c-Myb} binds with a preferred helical conformation, allowing the formation of native contacts and, in the last step, folding and binding of the C-terminal.
Study of the fluxes derived from the MSM shows the relevance of residue Leu302, not only in the final bound structure but also as the responsible for establishing the first contacts and serving as an anchoring point between \mbox{c-Myb} and KIX. 

%The role of Leu298 and Leu302 as the driving force for the interaction was originally described in the report on the NMR structure \cite{zor2004solution} and was fatherly supported by mutagenesis studies showing the abolition of binding upon their mutation to alanine \cite{giri2013structure}.

The model describes an overall induced-fit binding mechanism, as the complete folding of cMyb is only observed when native contacts have been formed. Conformational selection would only affect the first binding stage on residues 298 to 302 and not the whole length of the peptide, whereas the latter stages of binding follow an induced-fit mechanism.

Overall, our results provide a detailed mechanistic model for the binding of \mbox{c-Myb} to the primary interface of KIX, as well as showing the interaction with a secondary binding site, by using unbiased full-atom MD simulations and MSM analysis. The novel MD sampling approach used in this work, AdaptiveBandit, had a crucial role in resolving this type of folding and binding process. 
The method is implemented and available in the HTMD python package \cite{doerr2016htmd}. 
However, more algorithms can be derived within the same bandit framework. 
While here we choose the reward to be minus the free energy, other choices could optimize different costs, for example, improving the precision of the off-rate or optimizing sampling in the context of structure prediction. 

\section{Methods}\label{methods}

\subsection*{Molecular dynamics simulations}

In order to generate initial conformations for \mbox{c-Myb} (residues 291 to 315), we ran multiple parallel simulations. The peptide was solvated in a cubic water box of $64$ \AA\ side with a NaCl concentration of 0.05 $M$.
First, the peptide was simulated at \mbox{500 $K$} for \mbox{120 ns} to unfold the initial structure. 
Then, 200 systems were built by placing one random unstructured \mbox{c-Myb} conformation in conjunction with KIX in opposite corners of a 64 \AA\ side cubic water box with a NaCl concentration of 0.05 $M$, resulting in a final protein concentration of $\sim$3.2 mM. 

All systems were built using HTMD \cite{doerr2016htmd} and simulated with ACEMD \cite{harvey2009acemd}, CHARMM22* force field \cite{piana2011robust} and TIP3P water model \cite{jorgensen1983comparison}. 
A Langevin integrator was used with a damping constant of 0.1 ps\textsuperscript{-1}. The integration time step was set to 4 fs, with heavy hydrogen atoms (scaled up to four times the hydrogen mass) and holonomic constraints on all hydrogen-heavy atom bond terms. 
Electrostatics were computed using PME with a cutoff distance of 9 \AA\ and grid spacing of 1 \AA.
After energy minimization, equilibration for all systems was done in an NPT ensemble at 303 K, 
1 atm,
with heavy atoms constrained at 1 $kcal\ mol^{-1}$ \textit{\AA\textsuperscript{2}}. 
Energy minimization was run for 500 steps and equilibrated for 2 ns.

Production runs of 250 $ns$ were performed at 310 $K$ using the distributed computing project GPUGrid \cite{buch2010high}, following an adaptive sampling strategy. The final data set included 1,809 trajectories of 250 ns, resulting in an aggregated simulation time of $\sim$450 $\mu$s.
Additionally, a set of long MD runs were performed starting from bound structures.
Four models of the NMR-determined structure and four random bound conformations were selected and equilibrated as previously described.
A total of 8 long trajectories of $\sim2\ \mu s$ each were generated.

\subsection*{Markov state model analysis}
The projected space used for building the MSM included four different featurizations: all pair C$_\alpha$ + C$_\beta$ atoms distances between KIX and \mbox{c-Myb} to account for the interaction between the two proteins, self-distances between every C$_{\alpha}$ of \mbox{c-Myb} and its secondary structure, to monitor its conformation, and finally, RMSD to the bound structure. 
TICA was used at a lag time $\tau=20$ ns (implied timescales are shown in Supplementary Figure 1.a) for both the distance features and the secondary structure features, taking the 4 most relevant components from the distance features (both inter-distances and cMyb self-distances) and the 3 most relevant components from the secondary structure features.

The 8-dimensional projected data was discretized into 2,000 clusters using the mini-batch k-means algorithm \cite{pedregosa2011scikit}. 
The microstates defined in the MSM were coarse-grained into larger meta-stable macrostates by using PCCA++ \cite{roblitz2013fuzzy}. 
For the estimation of kinetic values, the original MSM was modified by creating an additional macrostate, considered as the \textit{bulk} state for all subsequent calculations to obtain the kinetics of binding. The bulk state was created by taking those microstates where the minimum  distance between KIX and cMyb was higher than a threshold. 
Error in kinetic measures was estimated by creating 50 independent MSMs using a random set containing 80\% of the simulation data.

To obtain the kinetic pathway of binding and folding, we  increased the number of macrostates in the MSM using PCCA++ again. Fluxes between macros were estimated using transition path theory \cite{weinan2006towards,noe2009constructing}. For the intra-macrostate flux analysis, we computed the mean helicity of cMyb for each microstate in it and clustered them into 4 main states which describe the peptide's grade of helix formation.
All analysis were performed with HTMD \cite{doerr2016htmd}. 

\subsection*{AdaptiveBandit sampling}
The multi-armed bandit problem is defined by $\langle\mathcal{A},\mathcal{R},\gamma\rangle $, where an action $a_t \in \mathcal{A}$  and $\mathcal{R}^a$ is a (stochastic) reward function. 
We choose $\gamma=0$ for totally discounted rewards. The optimal policy $\pi_a \sim \mathbb{P}[a]$ 
selects actions $a_t$ in order to maximize the cumulative future rewards. 
The construction of an optimal selection strategy requires handling the exploration-exploitation problem. AdaptiveBandit relies on the UCB1 algorithm \cite{auer2002using}, defining an upper confidence bound for each action-value estimate based on the number of times an action has been picked and the total amount of actions taken
\begin{equation}
    a_t = \argmax_{a\in\mathcal{A}}\left[{Q_t(a) + c\sqrt{\frac{\ln{t}}{N_t(a)}}}\right] ,
    \label{UCBeq}
\end{equation}
where $t$ denotes the total number of actions taken, 
$Q_t(a)=\mathbb{E}_\pi [ r \vert a ] $ is the action-value estimation, $N_t(a)$ 
is the number of times action $a$ has been selected (prior to time $t$) and $c$ is a constant controlling the degree of exploration. 
As for the reward definition, there are different choices depending on the objective, e.g. here, the interest is sampling the bound metastable state, hence, we rewarded actions based on the stability of conformations using MSM estimations of the free energy for each state
\begin{equation}
    \mathcal{R}_a = <k_{B}T\log(\mu(x))>_{(a,x_1,\dots, x_{\tau})},
    \label{reweq}
\end{equation}
where $\mu(x)$ is the equilibrium distribution estimated by the MSM with the currently available data and the average is performed over the frames in the trajectory starting from $a$. AdaptiveBandit uses the MSM discretized conformational space to define the action set and at each round acquires a random conformation from the selected states to respawn new simulations.
A more formal description of the bandit framework and AdaptiveBandit in the context of adaptive sampling as well as analysis in simpler, analytical potentials are available at \cite{perez2020adaptivebandit}. The AdaptiveBandit sampling algorithm is made available in the HTMD \cite{doerr2016htmd} Python package.

\subsection*{Adaptive Sampling parameters}
For both the AdaptiveBandit and the count Adaptive runs, the construction of MSMs at each epoch was done using the residue-residue contacts between KIX and \mbox{c-Myb} measured as the minimum contacts between residues at a threshold of 5 \AA, and the backbone dihedral angles of \mbox{c-Myb}. 
Time independent component analysis (TICA) \cite{perez2013identification} was used for dimensionality reduction using a lag time of $\tau=20$ frames and 
keeping the 3 first dimensions, which were later clustered with a k-centers algorithm. 
AdaptiveBandit was performed during 40 epochs with a $c$ value of 0.01.

\begin{acknowledgement}

The authors thank Kresten Lindorff-Larsen and Frank No\'{e} for their critical reading of the manuscript. The authors also thank volunteers at GPUGRID.net for contributing computational resources and Acellera for funding. G.D.F. acknowledges support from MINECO (Unidad de Excelencia Mar\'{i}a de Maeztu CEX2018-000782-M and BIO2017-82628-P) and FEDER. This project received funding from the European Union’s Horizon 2020 Research and Innovation Programme under Grant Agreement No. 823712 (CompBioMed2 Project).

\end{acknowledgement}

\section{Author Contributions}
P.H.N. and A.P. generated and analyzed the data; P.H.N., A.P. and G.D.F. wrote the paper; G.D.F. designed research.

\begin{suppinfo}

Supporting information is available free of charge.
\begin{itemize}
  \item Supporting information: Contains all supporting figures as well as the supporting table containing trajectory videos capturing the folding and binding of \mbox{c-Myb} with KIX domain.
\end{itemize}

\end{suppinfo}
\bibliography{achemso-demo}

\end{document}

\begin{table}
\centering
\caption{\textbf{Visualizations of representative trajectories showing the binding and folding of cMyb with the KIX domain.} The selected trajectories were reconstructed from different trajectory fragments spawned across different epochs in the adaptive sampling scheme. The seven trajectories listed represent the seven unique paths that sample conformations with RMSD lower than 4 \AA\  to the bound pose.  }
\begin{tabular}{llll}
\hline
\hline
Trajectory & Min RMSD (\AA) & Trajectory length [ns] & Video URL \\
\hline
\hline
Trajectory 1 & 3.0 & 323.1 & \url{https://youtu.be/XNCe88Yzxro} \\ 
\hline
Trajectory 2  & 3.1 & 631.7 &  \url{https://youtu.be/fZZji2cdaKI} \\ 
\hline
Trajectory 3 & 3.3 & 323.1 &  \url{https://youtu.be/6qYiyhsqtig} \\ 
\hline
Trajectory 4 & 3.5 & 706.5 &  \url{https://youtu.be/d8nvW8i_vHk} \\ 
\hline
Trajectory 5 & 3.7 & 521.7 &   \url{https://youtu.be/eAOnZcW3D5Y}\\ 
\hline
Trajectory 6 & 3.8 & 445.1 & \url{https://youtu.be/vltn2zURsfY} \\ 
\hline
Trajectory 7 & 4.0 & 323.1 & \url{https://youtu.be/Mn_wjOxmcvU} \\ 
% \hline
\hline
\end{tabular}
\label{tab:folding_movies}
\end{table}
\clearpage

%%% Each figure should be on its own page
\begin{figure}
\centering
\includegraphics[width=\linewidth, keepaspectratio]{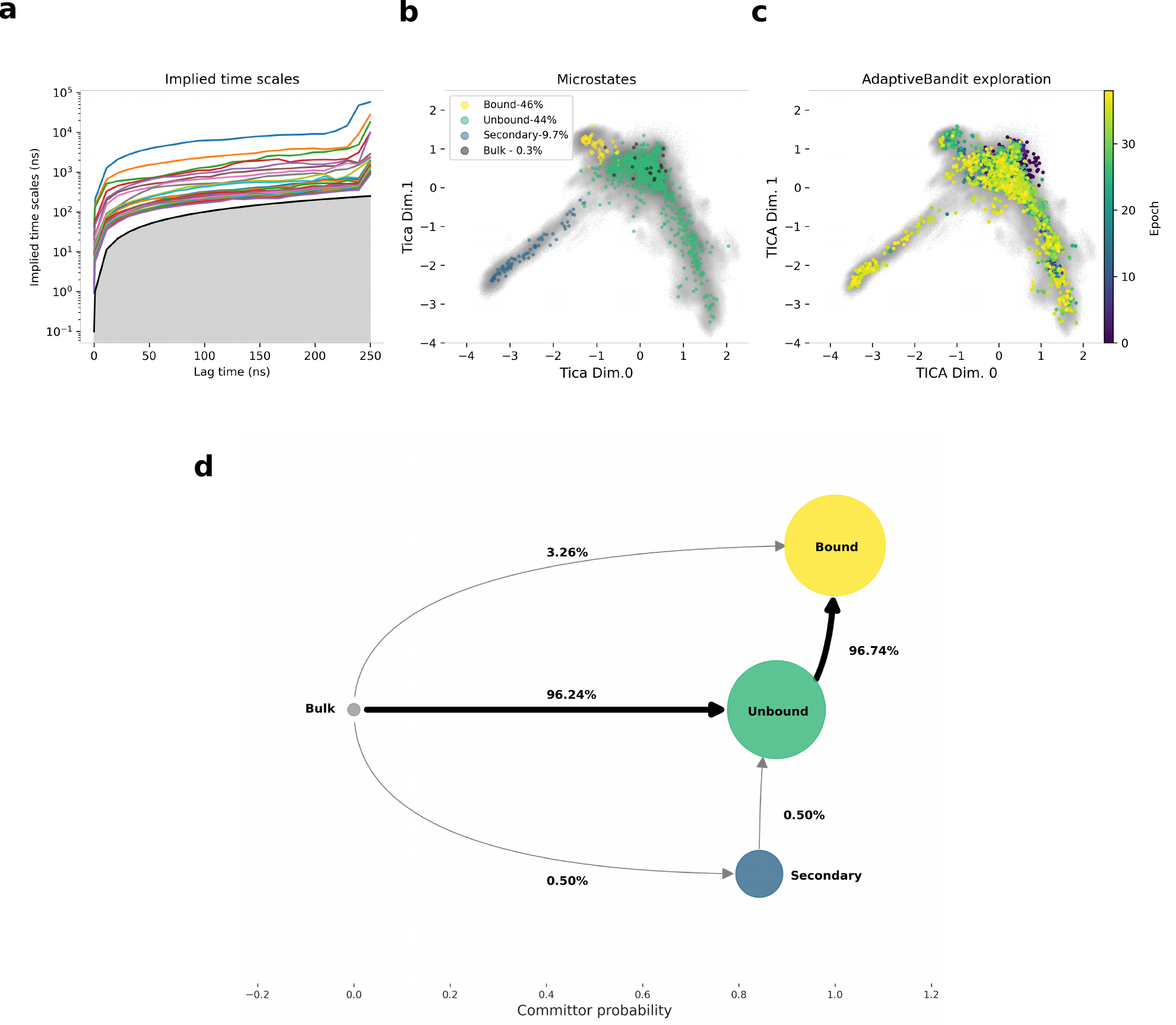}
\caption{\textbf{Markov state model summary} 
\textbf{a)} Implied time scales of the MD data.
\textbf{b)} Microstate distribution across the first two TICA dimensions. Each microstate is colored by its corresponding macrostate. The legend shows the population of each macrostate. 
\textbf{c)}  AdaptiveBandit exploration of the TICA space. Each colored point indicates a starting point selected by AdaptiveBandit to respawn a new trajectory. The color indicates the epoch number. In grey, the area covered by the projected simulation data without clustering, both in b) and c). \textbf{d)} Flux pathway from \textit{bulk} to \textit{bound}. Nodes are placed according to the committor probability. The \textit{y axis} is manually set for better visualization of the graph. Node size is proportional to the equilibrium distribution. Node color corresponds to macrostate assignment as in b). The flux percentage is shown near each arrow. The main pathway is indicated with black, thicker arrows. 
}
\label{fig:si_tica_explore}
\end{figure}

\clearpage

\begin{figure}
\centering
\includegraphics[width=13cm,keepaspectratio]{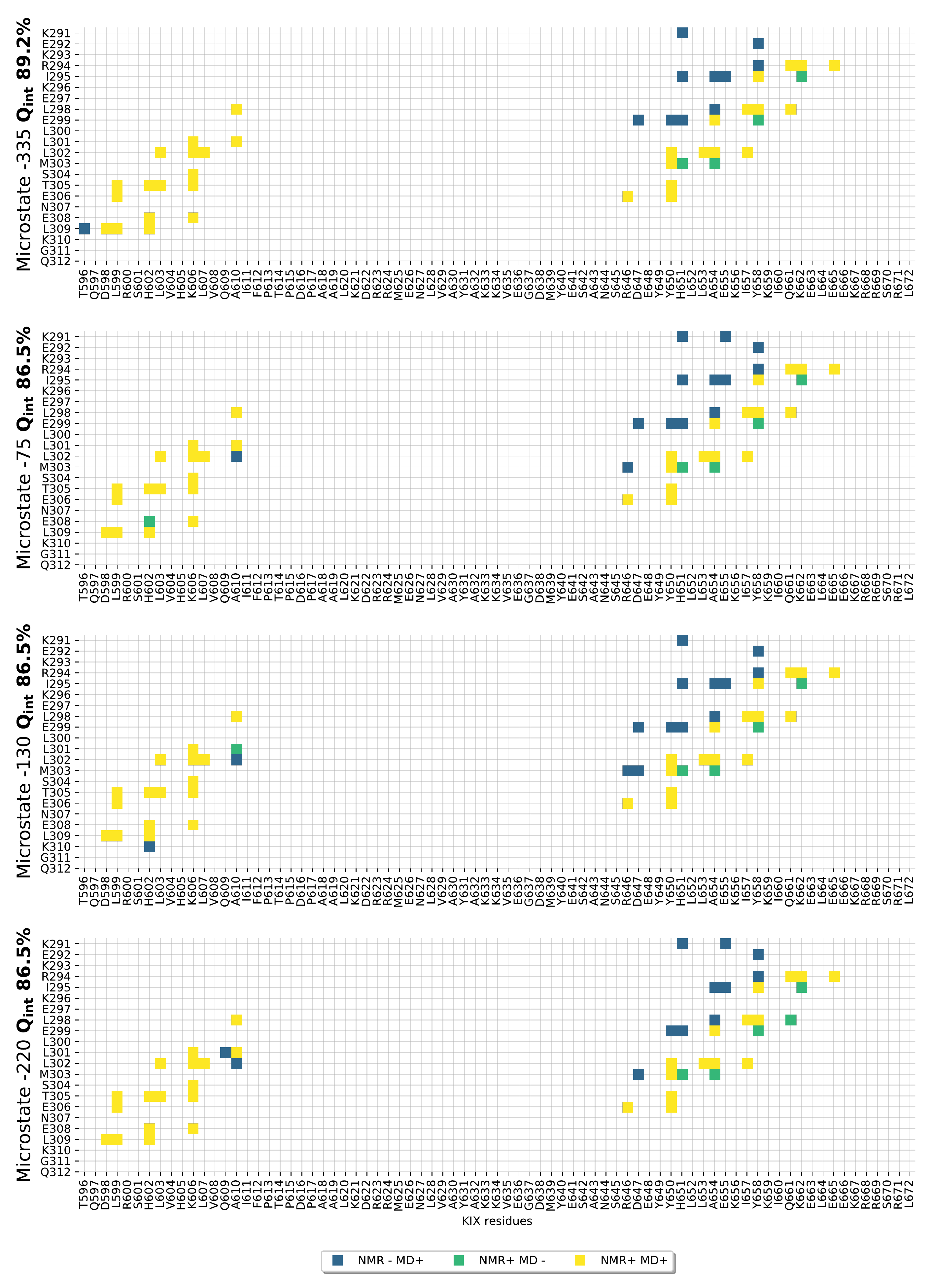}
\caption{\textbf{Maximum  $\boldsymbol{Q_{int}}$ microstates contact fingerprint.} Profile of contacts established between c-Myb and KIX in microstates with maximum fraction of native binding contacts $Q_{int}$. 
Blue color represents contacts present in the state but not in the original NMR conformation, green indicates native contacts not found in the MSM state and yellow squares represent a match on that contact, found in both the NMR model and MD microstate. A contact is considered present in a microstate when it appears in at least 50\% of the conformations in that state.
}
\label{fig:si_contacts}
\end{figure}

\clearpage

\begin{figure}
\centering
\includegraphics[width=\linewidth, keepaspectratio]{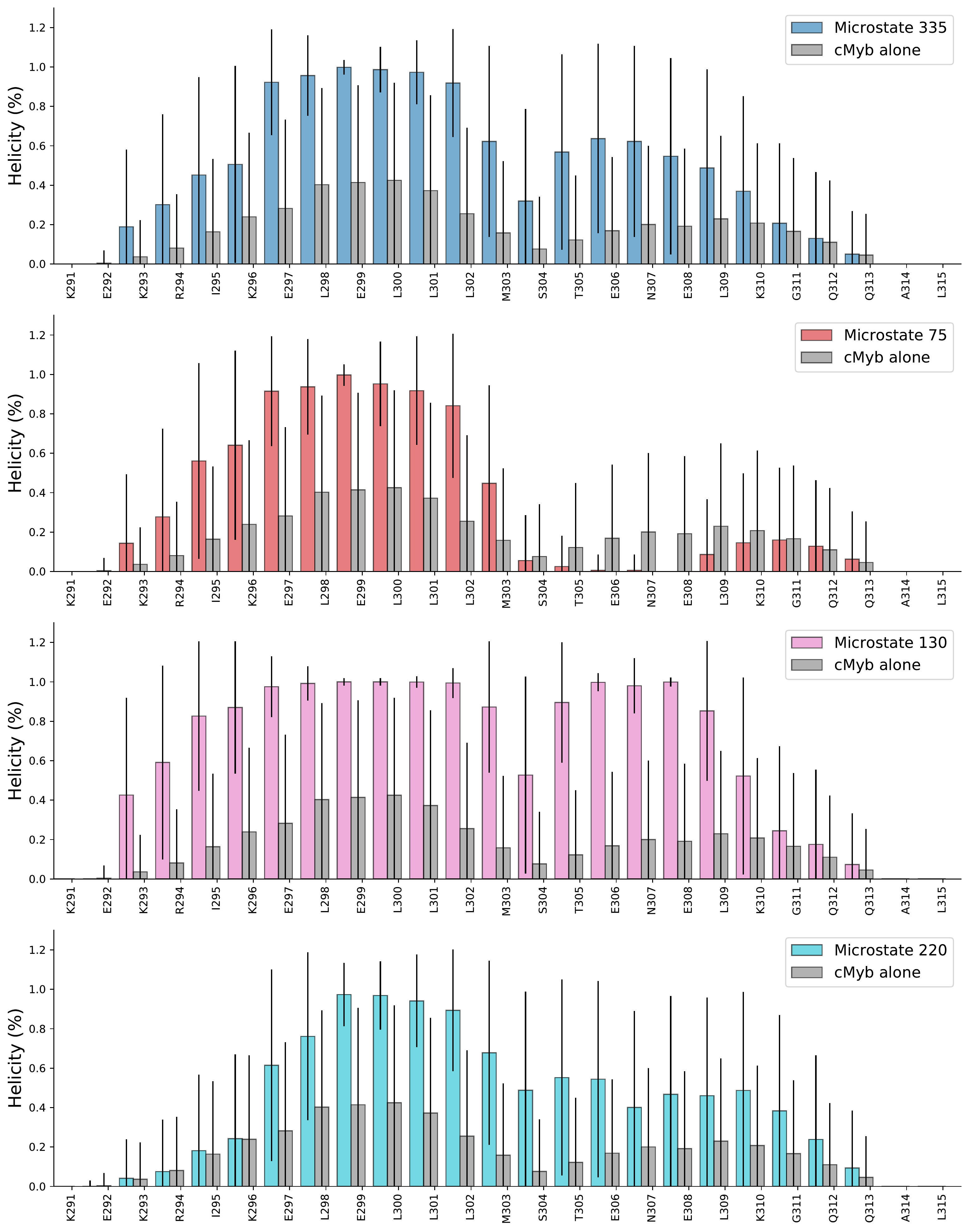}
\caption{\textbf{c-Myb helicity.} Comparison of the by-residue helicity fraction of c-Myb between the four microstates with maximum $Q_{int}$. The helicity profile for the peptide in isolation is depicted in grey.
}
\label{fig:si_helicity}
\end{figure}

\clearpage

\begin{figure}
\includegraphics[width=\linewidth, keepaspectratio]{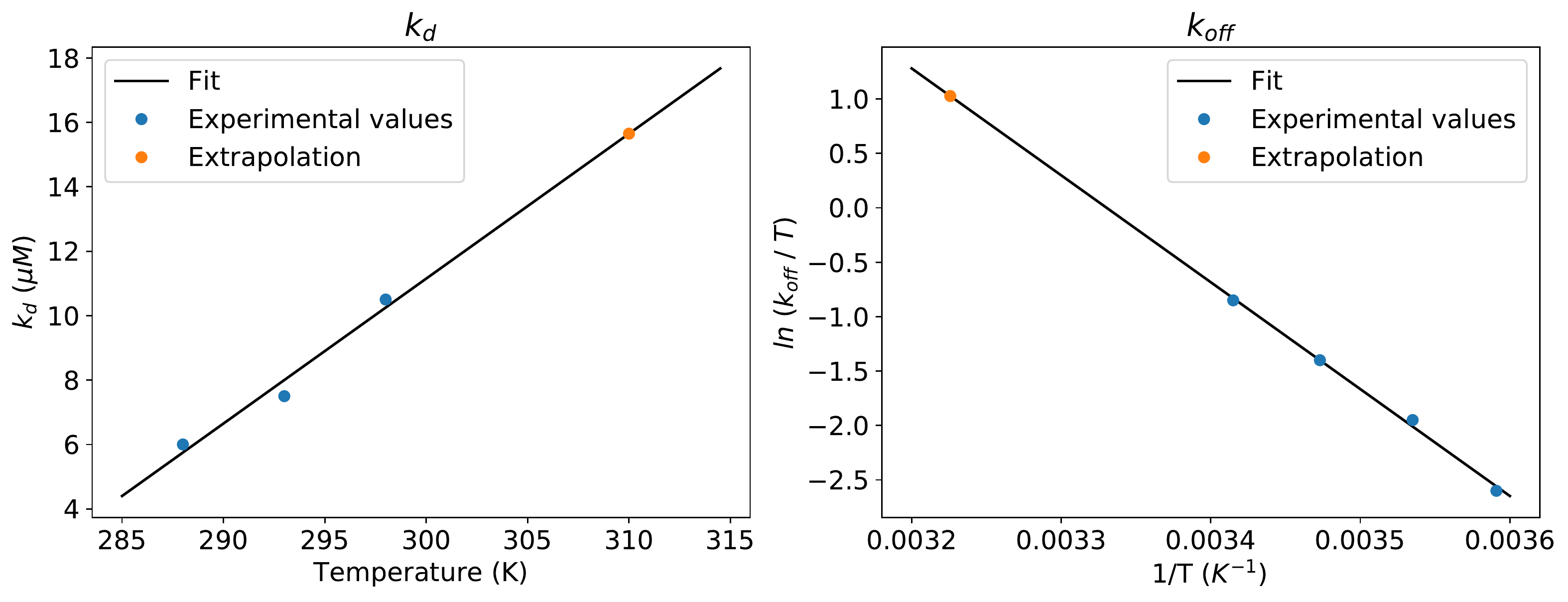}
\caption{Extrapolations of \textbf{a)} $k_d$ and \textbf{b)} $k_{off}$ values from experimental data. 
}
\label{fig:si_extrapolation}
\end{figure}

\clearpage

\begin{figure}
\centering
\includegraphics[width=11.8cm, keepaspectratio]{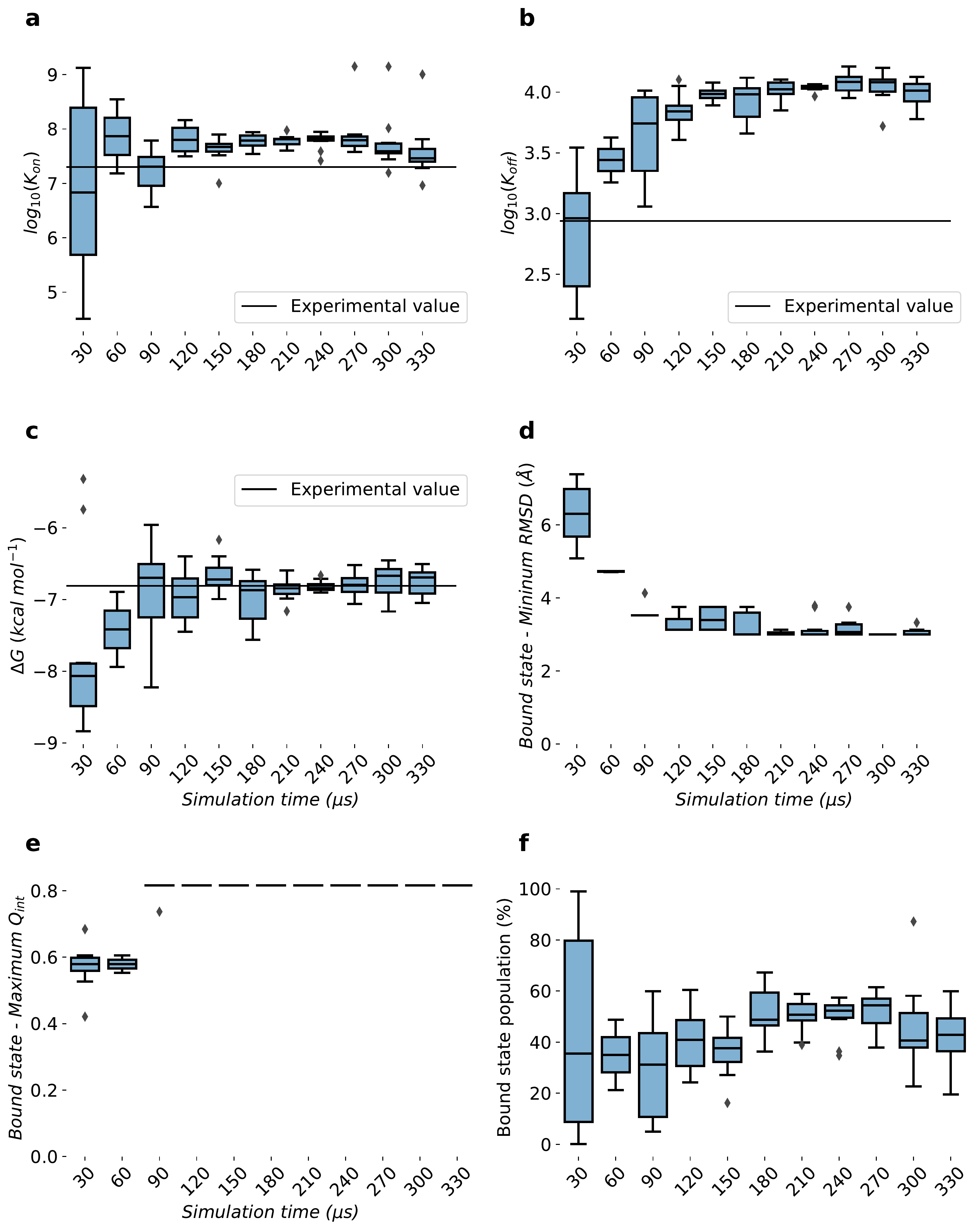}
\caption{\textbf{Statistics convergence across the MD run} of \textbf{a)} k\textsubscript{on}, \textbf{b)} k\textsubscript{off}, \textbf{c)} free energy and \textbf{d)} microstate minimum RMSD, \textbf{e)} bound state maximum $Q_{int}$ and \textbf{f)} bound state population, computed by the MSM. Each data point was calculated by building 10 different MSMs, bootstrapping 80\% of the trajectories each time.
}
\label{fig:si_epochs}
\end{figure}

\clearpage

\begin{figure}
\centering
\includegraphics[width=11.8cm, keepaspectratio]{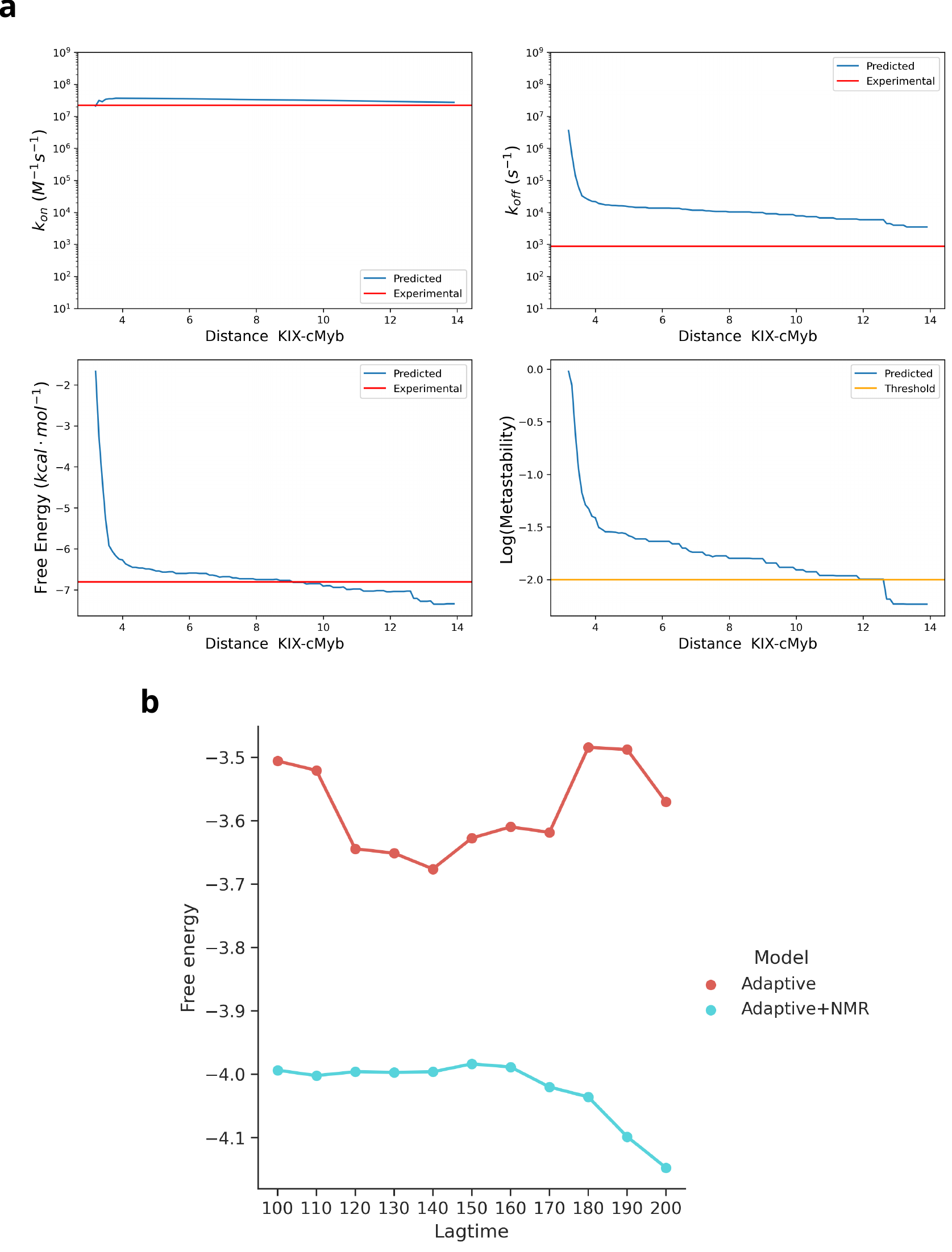}
\caption{\textbf{a)} Bulk state variability of k\textsubscript{on}, k\textsubscript{off}, free energy and metastability term, depending on the maximum distance threshold between KIX and cMyb used. The metastability term is defined as the bulk state self-transition probability (named for plot simplicity). The blue line shows the variable estimate, the red line shows the reference experimental value and the yellow line shows the defined threshold for deciding whether the bulk state is stable enough or not. \textbf{b)} Free energy estimates at different MSM lag times for the AdaptiveBandit simulations alone and together with the long trajectories starting from the NMR structures. The models were performed without defining a bulk state. 
}
\label{fig:si_bulk}
\end{figure}

\clearpage

\begin{figure}
\centering
\includegraphics[width=\linewidth, keepaspectratio]{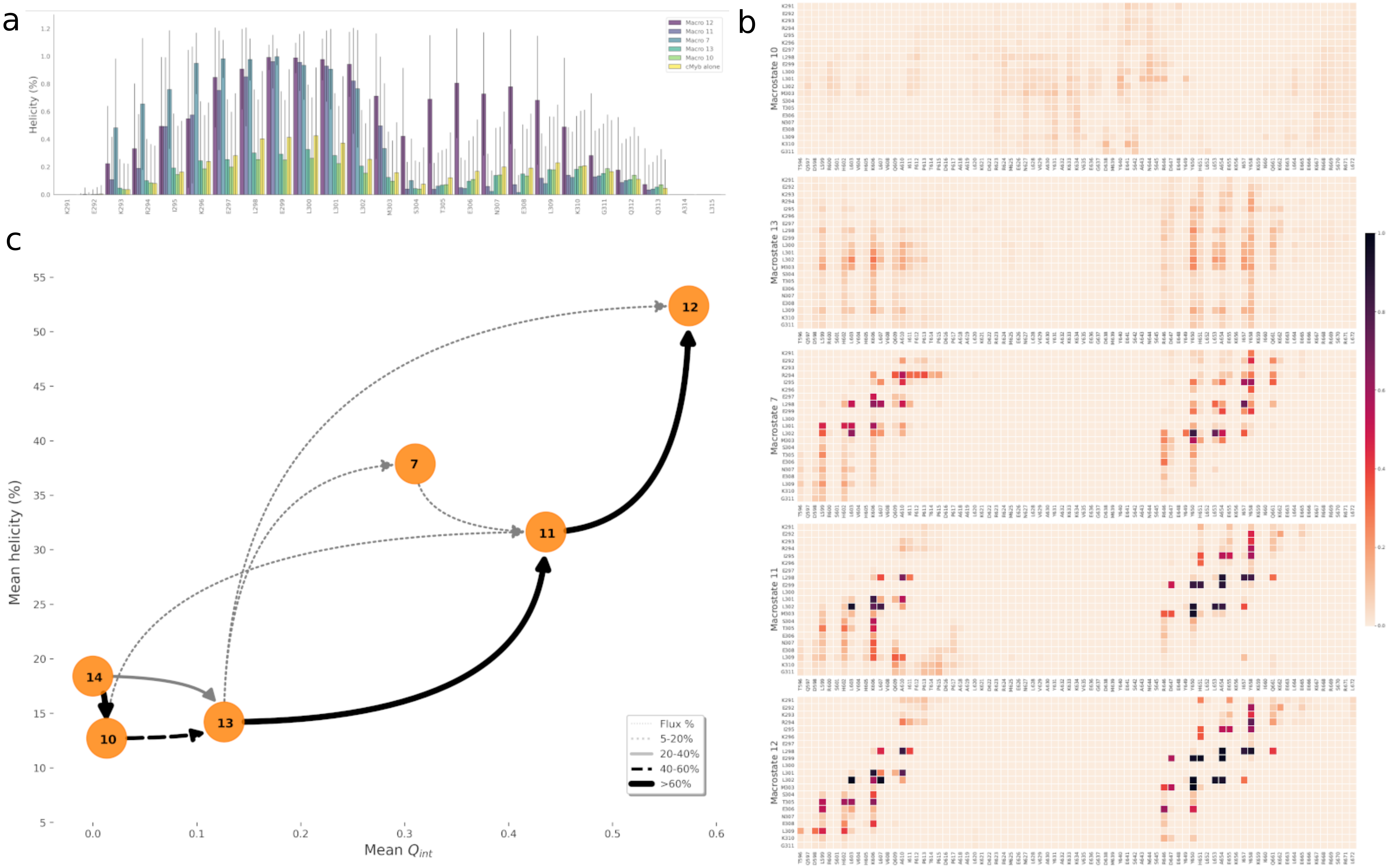}
\caption{\textbf{Complete binding process of c-Myb to KIX.} Structural analysis of the states involved in the main binding flux pathway on the 15 macrostate MSM.
\textbf{a)} Mean helicity per residue and \textbf{b)} mean contacts profile is shown for the macrostates present in the binding process. \textbf{c)} Main pathways leading from Macrostate 14 (\textit{Bulk}) to Macrostate 12 (\textit{Bound}). Nodes are placed according to the fraction of native contacts $Q_{int}$ with respect to the NMR model on the \textit{x axis}, and mean helicity on the \textit{y axis}.
Arrows represent the connection between macrostates, and their color, thickness and trace the percentage of the total flux traversing them.}
\label{fig:si_binding}
\end{figure}

\clearpage

\begin{figure}
\centering
\includegraphics[width=11.8cm, keepaspectratio]{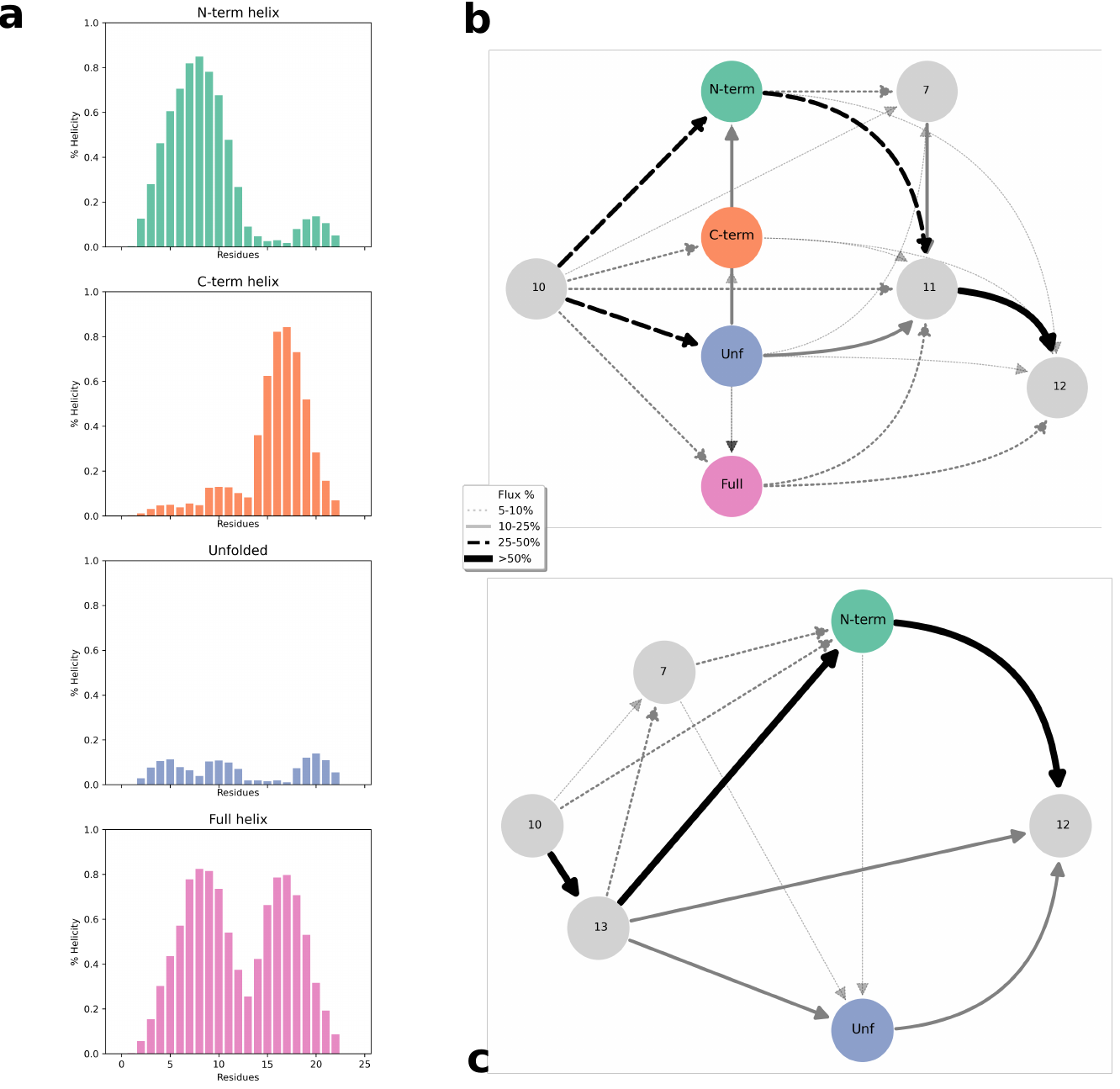}
\caption{\textbf{Detailed intra-macrostate flux analysis. a)} Cluster centers corresponding to the four main cMyb folded states: Full helix, Unfolded, N-terminal helix, C-terminal helix. These centers were computed using the mean helicity of all microstates. The centers are displayed with bar plots showing the helicity per residue. \textbf{b,c)} Flux pathways across clustered macrostates 13 (b) and 11 (c). The selected macrostates were clustered using the four centers defined in a), and the flux was recomputed using these newly defined clusters. Node colours and names indicate which cluster center from a) they correspond. Node positioning was manually set for visualization purposes. Arrows represent the connection between macrostates/clusters. Their color, thickness, and trace represent the percentage of the total flux traversing them.
}
\label{fig:si_helixclu}
\end{figure}

\clearpage

\begin{figure}

\includegraphics[width=\linewidth, keepaspectratio]{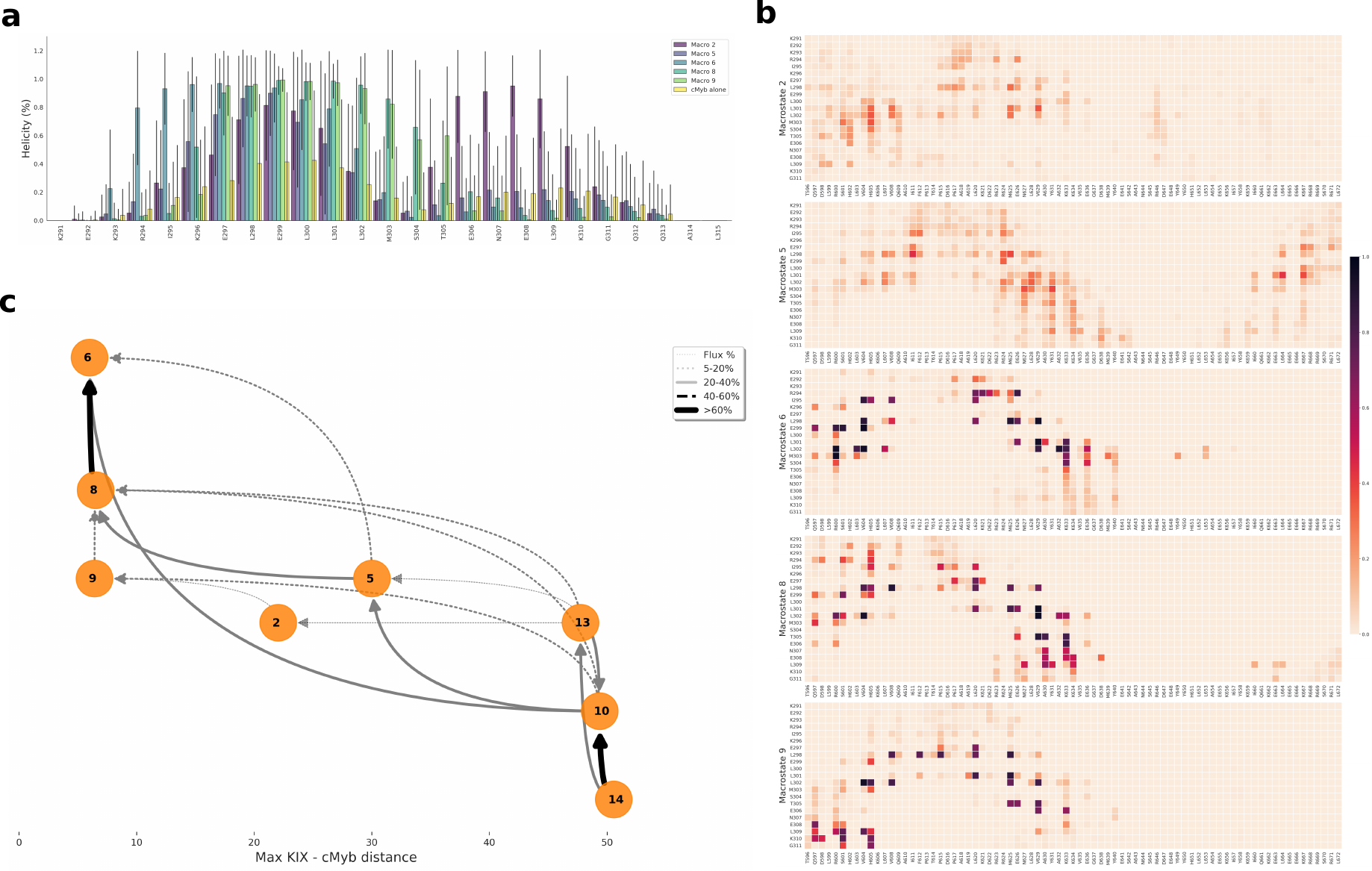}
\caption{\textbf{Secondary binding path of KIX and c-Myb.} Study of the states involved in the secondary binding pathway, using the 15 macrostate MSM.
For selected macrostates the \textbf{a)} mean helicity  and \textbf{b)} KIX---c-Myb contacts profile is shown.  Contact and helicity data for macrostates 10 and 13 are shown in FigS7. \textbf{c)} Main pathways leading from Macrostate 14 (\textit{Bulk}) to Macrostate 6 (\textit{Secondary}). Nodes are placed according to the maximum distance between KIX and cMyb of each state on the \textit{x axis}. The \textit{y axis} is manually set for better visualization of the graph. Arrows represent the connection between macrostates, and their color, thickness, and trace the percentage of the total flux traversing them.
}
\label{fig:si_pathway2}
\end{figure}